\documentclass{article}

\usepackage{arxiv}
\UseRawInputEncoding 
\usepackage[utf8]{inputenc} % allow utf-8 input
\usepackage[T1]{fontenc}    % use 8-bit T1 fonts
\usepackage[spanish]{babel}
\usepackage{hyperref}       % hyperlinks
\usepackage{url}            % simple URL typesetting
\usepackage{booktabs}       % professional-quality tables
\usepackage{amsfonts}       % blackboard math symbols
\usepackage{nicefrac}       % compact symbols for 1/2, etc.
\usepackage{microtype}      % microtypography
\usepackage{lipsum}
\usepackage{graphicx}
\usepackage{float}
\usepackage{amsmath}
\usepackage{xcolor}
\usepackage{url}
\graphicspath{ {./images/} }

\title{Overview of processing techniques for surface electromyography signals}

\author{
  Manjarres-Triana Alejandra \\
  Bioengineering Program\\
  Universidad El Bosque\\
  Bogotá, Colombia\\
  \texttt{mmanjarrest@unbosque.edu.co} \\
   \And
  Acevedo-Serna Juan \\
  Bioengineering Program\\
  Universidad El Bosque\\
  Bogotá, Colombia\\
  \texttt{jacevedos@unbosque.edu.co} \\
  \And
  Ramírez-Duque Andrés A. \\ 
  School of Computing Science \\
  University of Glasgow \\
  Glasgow, UK. \\
  \texttt{Andres.Ramirez-Duque@glasgow.ac.uk} \\
  \And
  Jiménez Mario F.\\
  School of Engineering, Science and Technology \\
  Universidad del Rosario \\
  Bogotá, Colombia \\ 
  \texttt{mariof.jimenez@urosario.edu.co} \\
  \And
  Pulido-Herrera Edith \\
  Bioengineering Program\\
  Universidad El Bosque\\
  Bogotá, Colombia\\
  \texttt{epulidoh@unbosque.edu.co} \\
  \And
  Villarejo Mayor John J. \\ 
  Postgraduation Program of Physical Education \\
  Federal University of Parana \\
  Curitiba, Brazil \\
  \texttt{jvimayor@gmail.com} \\
}

\begin{document}
\maketitle
\begin{abstract}
Surface electromyography (sEMG) is a technology to assess muscle activation, which is an important component in applications related to diagnosis, treatment, progression assessment, and rehabilitation of specific individuals' conditions. Recently, sEMG potential has been shown, since it can be used in a non-invasive manner; nevertheless, it requires careful signal analysis to support health professionals reliably. This paper briefly described the basic concepts involved in the sEMG, such as the physiology of the muscles, the data acquisition, the signal processing techniques, and classification methods that may be used to identify disorders or signs of abnormalities according to muscular patterns. Specifically, classification methods encompass digital signal processing techniques and machine learning with high potential in the field. We hope that this work serves as an introduction to researchers interested in this field.
\end{abstract}

% keywords can be removed

\section{Introduction}
\label{sec:introduction}

Electromyography allows for assessing neuromuscular activity and muscle activation, making it appealing for applications in sports, rehabilitation, or biofeedback, among others \cite{Merletti_2010}. Surface Electromyography (sEMG) is a non-invasive technique to measure muscle activation, however, given its complexity, it has hardly been utilized in clinical practice and rehabilitation \cite{Pilkar_2020}. On the other hand, studies have shown the potential of the sEMG to identify musculoskeletal disorders through different sets of muscles by generating discriminant electrical patterns \cite{Mordhorst_2015}. In order to leverage the sEMG to overcome its barriers, clinicians and researchers require training and knowledge of good practices to use this technology \cite{Pilkar_2020, Merletti_2020}. In addition, for an accurate diagnosis, it is required a good characterization of the muscle patterns according to the disease's patient \cite{Yousefi_2014}. In this sense, sEMG allows conducting of quantitative evaluation of muscle patterns, allowing discriminating intrinsic relationships from the information provided that facilitate the patient status muscle \cite{Drost_2013}.

In this work, we provide an overview of the basic topics associated with sEMG processing data to conduct quantitative analysis. At first, we describe concepts of the neuromuscular system used in the sEMG systems (see section \ref{sec:Electromyography}), and the EMG acquisition basic aspects are described in section \ref{sec:Acquisition}. Preprocessing techniques of EMG signals, such as the filters, epochs, and frequency spectral density, among others, are described in section \ref{sec:preprocessing}. Next, relevant features are presented (see section \ref{sec:Feature}) to be used in classification techniques including statistical methods and machine learning, and lastly, some examples are presented. By covering all these topics, we aim to provide a general overview to interested researchers in the topic.

\subsection{Origin of electromyographic signals}
\label{sec:Origin_EMG}

\begin{figure}[h!]
    \centering
    \includegraphics[width=0.5\textwidth]{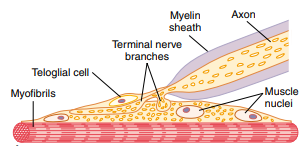}
    \caption{Neuromuscular system, the longitudinal section between the single axon terminal and the muscle fiber membrane \cite{Guyton_2006}.}
    \label{fig:neurosystem}
\end{figure}

In the neuromuscular system, the central nervous system (CNS) controls muscle contractions through nerve signals from the brain via the spinal cord. Final motor commands are related to the Motor Unit (MU), the smallest functional unit of the nervous system. A MU is composed of muscle fibers innervated by a single motor neuron (see Fig. \ref{fig:neurosystem}); therefore, when recruited results in a muscular contraction.

Producing precise or gross motion patterns is related to a few or a high number of muscle fibers innervated. The number of MUs per muscle in humans can range from around $100$ for a hand muscle to $1000$ or more for large limb muscles. MUs have also been shown to vary greatly in force-generating ability, with a one hundred times difference or more in force of contraction. Three types of motor units are defined based on physiological properties such as conduction speed and muscle fatigue: (1) Fast-twitch, fatigable (FF or type IIb); (2) fast-twitch, fatigue-resistant (FR or type IIa); and (3) slow-twitch (S or type I), which is more resistant to fatigue \cite{Guyton_2006}.

\begin{figure}[h!]
    \centering
    \includegraphics[width=0.6\textwidth]{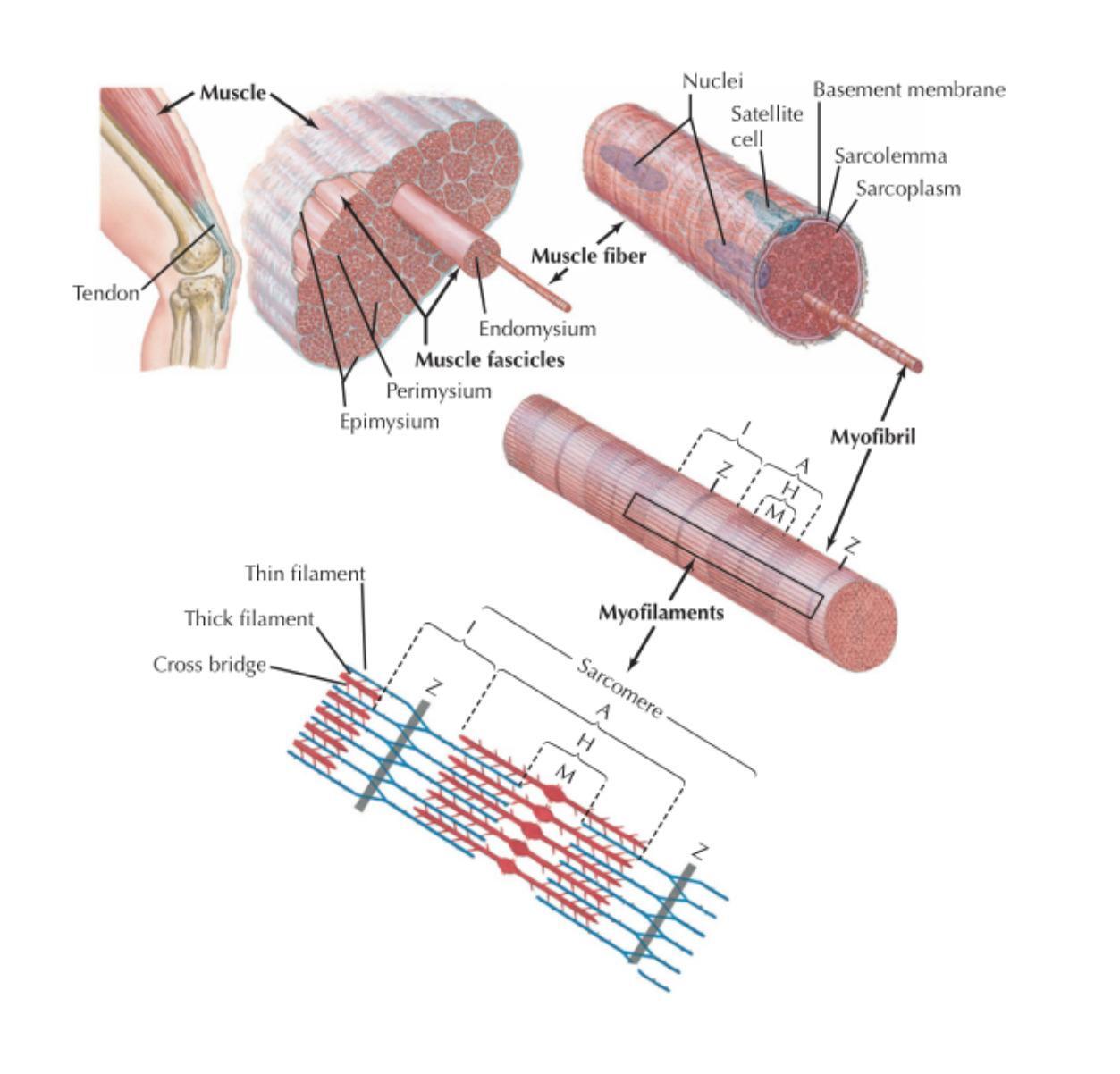}
    \caption{Skeletal muscle parts, the muscles that connect to your bones and allow you to perform a wide range of movements and functions \cite{Thompson_2010}}
    \label{fig:Skeletal}
\end{figure}

The functional unit of skeletal muscle contraction is the sarcomere, which is located between two Z disks (see Fig. \ref{fig:Skeletal}). Skeletal muscle is made up of bundles, fibers, and myofibrils; myofibrils are divided into two types: actin (thin filament) and myosin (thick filament) \cite{Guyton_2006}.

\begin{figure} [h!]
    \centering
    \includegraphics[width=0.4\textwidth]{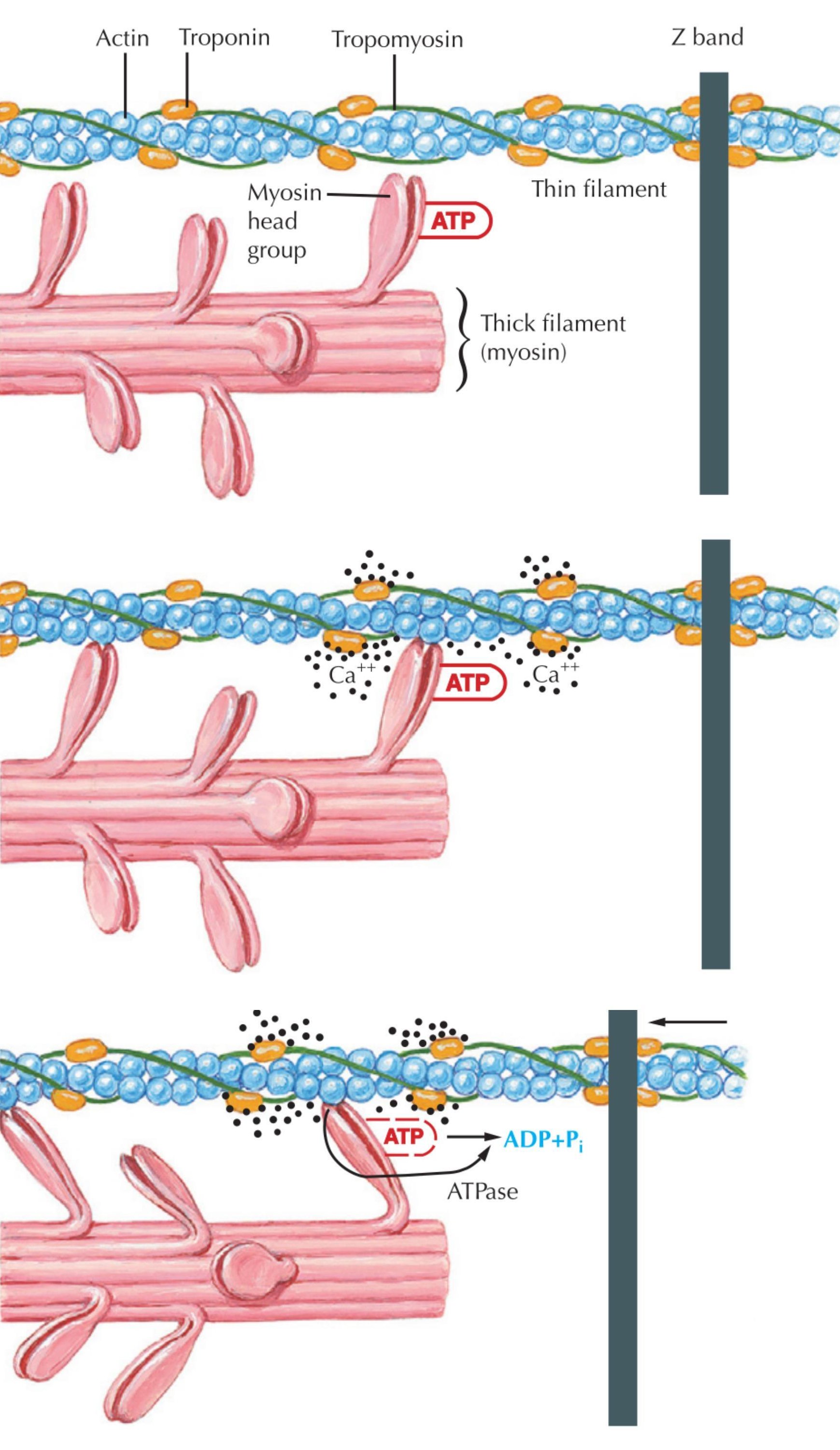}
    \caption{Muscle contraction, at the molecular level, muscle contraction is defined by myosin molecules pulling actin filaments \cite{Thompson_2010}.}
    \label{fig:musclecontract}
\end{figure}

During muscle contraction, the myosin head must be activated before the step-by-step mechanism of contraction begins. This occurs when adenosine triphosphate (ATP) binds to the myosin head and undergoes hydrolysis, leaving adenosine diphosphate (ADP) and inorganic phosphate. The energy released from ATP hydrolysis activates the myosin head. The activated myosin head binds to actin and inorganic phosphate is released (the binding becomes stronger), then ADP is liberated and the myosin head is displaced, causing the actin filament to move toward the midline. Another ATP comes and binds to the myosin head (the binding is weakened) and separates from actin, then, the myosin head is activated again (see Fig. \ref{fig:musclecontract}). The mechanism ends when calcium is pumped into the sarcoplasmic reticulum and the tropomyosin returns to its original place, so the myosin head no longer has a place to join \cite{Guyton_2006}.

\subsection{Acquisition of electromyographic signals}
\label{sec:Acquisition}

The EMG signal is a representation of the electrical activity of the active muscle fibers during a contraction. The tissues act as spatial low-pass filters in the distribution of potential. Its detection can be intramuscular or superficial. The surface EMG (sEMG) technique is a non-invasive manner to capture myoelectric activity since it uses electrodes placed on the surface of the skin, in contrast with intramuscular EMG (iEMG) which uses invasive needles. For sEMG, the tissue acts as a conductive volume between the muscles and the electrodes. Therefore, tissue properties influence signal characteristics in terms of frequency content and distance beyond which the signal cannot be detected \cite{Merletti_2004}.

Regarding electrical properties, sEMG signals have an amplitude of $0.05-10$ mV and a frequency range of $2-500$ Hz \cite{Guerrero_2014}. SEMG signals are random, non-stationary, and nonlinear (no linear relationship between muscle activity and sEMG signal pattern) and are not generated by periodic phenomena. However, they are susceptible to analysis with linear techniques considering small time windows ($<500$ ms) where they can be considered stationary \cite{Merletti_2004}.\cite{Mohr_2018}.

\begin{figure}[h!]
    \centering
    \includegraphics[width=0.3\textwidth]{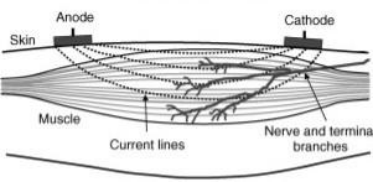}
    \caption{Bipolar EMG measurements, two adjacent electrodes are applied between innervation zone and tendon \cite{Merletti_2016}.}
    \label{ref:bipolar}
\end{figure}

\subsubsection{Factors influencing the EMG signal}
\label{sec:actors}

The detection of EMG signals can be affected by different factors that alter their shape and characteristics, which may affect signal processing and analysis.

As the surface electrodes capture the electric field generated by the depolarization of the muscle fiber, electrical conduction is affected by the tissue characteristics, since a great amount of adipose tissue influences the signals amplitude \cite{Konard_2005}.

Moreover, nearby muscles influence the EMG signal from a target muscle known as crosstalk. Then, tight arrangements within muscle groups need special care. Also, the signal produced by the depolarization of the muscle fibers of the heart infers with the recording of the EMG signal, mainly when analyzing muscles of the upper part of the trunk \cite{Konard_2005}.

Changes in geometry between the muscle and the electrodes are also a factor to be taken into account. In dynamic muscle evaluations, elongation of the skin is generally present, which may increase the distance between the source of the signal and the detection site \cite{Konard_2005}.

The environment artifacts contaminate the EMG signals, being the interference of the electrical line noise at $50$ or $60$ Hz is the most recurrent due to incorrect grounding of other external devices \cite{Konard_2005}.

\section{Preprocessing}
\label{sec:preprocessing}

Preprocessing techniques are useful tools to compress and highlight relevant information from biological signals, to increase their correlation. This section contains filters, epochs, frequency spectral density, and normalization of the maximum voluntary contractions of each subject \cite{Melo_2012}.

\subsection{Filters and offset removal}

Digital filters are usually applied to remove artifacts. The Butterworth filter has been widely used due to that it can maintain the frequency response as flat as possible in the passband. Also, this filter maintains the shape for higher orders\cite{Butterworth_1930}. In electromyography, it is recommended to use a 4th-order Butterworth filter between $15-400$ Hz, which has the characteristic of transmitting a range of frequencies and rejecting two frequency bands \cite{Quiroz_207}. The noise frequency of the electrical network can be attenuated by implementing an adaptive filter at $60$ Hz and its harmonics because there is less impact on the response of the filter due to variations in its components. It is also important to eliminate the offset of each signal. This is done by subtracting its respective mean since if the signal is centered at the origin, the offset level is said to be $0$. If it is displaced upwards, the offset is positive, while if it is displaced downwards, it will be negative \cite{Salgado_2015}.

\subsection{Epochs and overlap}
\label{sec:Epochs} 

\begin{figure}[h!]
    \centering
    \includegraphics[width=0.5\textwidth]{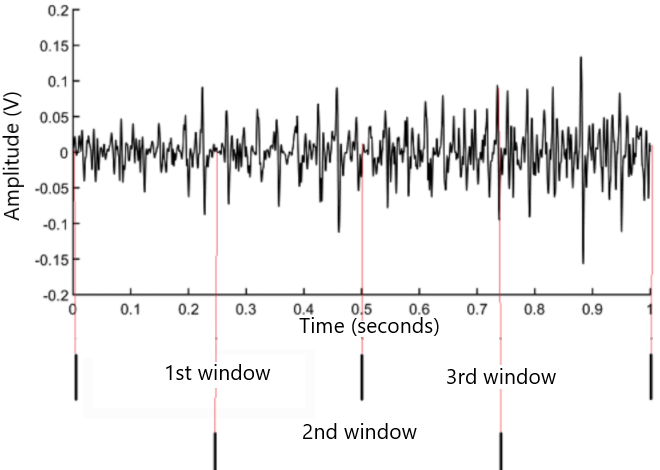}
    \caption{$500$ ms windows with $50\%$ overlap.}
    \label{fig:fig5}
\end{figure}

It is important to select the muscle electromyography signal when the exercise is being performed, therefore a segmentation of the signal must be carried out where only the segment is taken into account during the test. The electromyography signal is not stationary, therefore there are biases related to the spectral and temporal estimation. Therefore, it is necessary to use epochs (windows) to reduce the bias. With epochs of $500$ ms the signal can be considered as stationary \cite{Merletti_2004}. To reduce information loss, it is necessary to use windows with $50\%$ overlap (see Figure \ref{fig:fig5}).

\subsection{Power spectral density}
\label{sec:Power_spectral}

During myoelectric fatigue, there is a reduction in the conduction velocity of the muscle fibers, which results in compression and a shift toward low frequencies. For this reason, it is more convenient to characterize fatigue using the different frequency parameters of the sEMG signal through the power spectral density \cite{{Correa_2016}}. 

The Welch periodogram method is an approach to determine the power spectral density of a signal at different frequencies, which uses periodogram spectrum estimations. These are obtained from signal conversion from the time domain to frequency domain \cite{{Welch_1967}}. The periodogram or sample spectrum is calculated directly from N samples of the electromyographic signal segment \cite{Mañanas_1999}.

\subsection{Normalization by Maximum Voluntary Contraction (MVC)}
\label{sec:normalization}

Normalization methods remove the influence of the log condition, whereas data are rescaled according to the maximum voluntary contraction. These methods also allow to carry out of a quantitative comparison of EMG signals between subjects \cite{Konard_2005}.

\subsection{Maximum voluntary contraction}
\label{sec:Maximum}

To determine the maximum voluntary contraction (MVC), static resistance exercises are performed with positions that allow maximum contraction. These tests must be carried out separately for each muscle investigated; if necessary, repeat the exercises three times for five seconds. Subsequently, the average of each MVC must be calculated, which represents the maximum amplitude that the muscle can have during a contraction \cite{Konard_2005}; importantly, this signal must be rectified and smoothed.

\subsection{Signal smoothing }
\label{sec:Signal} 

Signal smoothing allows evaluating only the signal that comes from muscle contraction, without considering contamination by mechanical movements or electromagnetic signals. The signal is smoothed by at first rectifying negative values, which are transformed into positive values. Then, with the rectified signal, smoothing is calculated, which creates a linear envelope of the signal. For smoothing there are different methods such as the moving average or to calculate the $RMS$ \cite{Altimari_2012}.

\section{Feature Extraction}
\label{sec:Feature}

In order to characterize muscular patterns associated with specific neuromuscular or musculoskeletal disorders, feature extraction is essential for a good classification process based on EMG signals \cite{Phinyomark_2011}. In this section, we present feature extraction associated with the fatigue and coactivation indexes.

\subsection{Fatigue index}
\label{sec:Fatigue}

\begin{figure}[h!]
    \centering
    \includegraphics[width=0.5\textwidth]{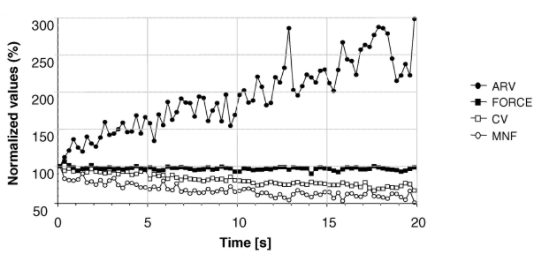}
    \caption{Parameters for muscle fatigue detection, where $ARV$ is the average rectified value, $CV$ the conduction velocity and $MNF$ the mean frequency, \cite{Merletti_2016}.}
    \label{fig:fig6}
\end{figure}

Myoelectric fatigue is the inability to continue generating a given level of force or exercise intensity \cite{Gomez_2009}, this is associated with decreased sensitivity and release of calcium ions \cite{Macintosh_2002}. Myoelectric fatigue is detected from different indexes in both the time and frequency domains, for example, motor action potential conduction velocity, root mean square, zero crossings, mean, and median of the frequency spectrum. Figure \ref{fig:fig6} shows an example of the graph of myoelectric fatigue detected from different indices in both the time and frequency domains (MNF, ARV and CV), observed during an isometric contraction of the anterior scalene muscle contracting the $2\%$ of maximal voluntary contraction \cite{Merletti_2016}.

EMG signals from moving muscles are expressed as time-bound waves. These waves can be used to determine time domain variables, such as the mean rectified value and zero crossings, and frequency domain variables, such as the mean and median \cite{Kim_2014}.

\subsubsection{Root Mean Square (RMS)}
\label{sec:rms}

The amplitude of the EMG signal is stochastic (random) with a Gaussian distribution that is related to the constant force \cite{Quinayas_2015} that varies from $0$ to $10$ mV (peak to peak). Different parameters are commonly used to measure the amplitude and find information on the energy of the time of the analyzed signal, such as the root mean square ($RMS$); which represents the square root of the average power of the EMG signal during a given period of time (see equation \ref{eq:rms}).

\begin{equation}
RMS = \sqrt{\frac{\sum_{n=0}^{N}EMG_{n}^{2}}{N}} 
\label{eq:rms}
\end{equation}

Where $RMS$ is the root mean square, $EMGn$ is the specific signal of a rectified channel, and $N$ is the number of samples.

\subsubsection{Median Rectified Value (ARV)}
\label{sec:ARV}

ARV is used to estimate the variation in amplitude of the EMG signal, whose coefficient of variation is lower than the $RMS$:

\begin{equation}
ARV= \frac{\sum_{i=1}^{N}|x_{i}|}{N}
\end{equation}

Where $ARV$ is the average rectified value, $X$ is the specific signal of a rectified channel and $N$ is the number of samples \cite{Mora_2013}

\subsubsection{Zero Crossings}
\label{sec:Zero}

Zero crossing measures the signal frequency, determining the number of times it crosses zero. A threshold is required to reduce the number of noise-induced zero crossings (see equation \ref{eq:zc}) and is selected per the signal voltage \cite{Romo_2007}.

\begin{equation}
ZC=EMG(n)>0 \quad \& \quad EMG(n+1)<0 
\quad \| \quad 
 EMG(n)<0 \quad \& \quad EMG(n+1)>0
 \label{eq:zc}
\end{equation}

Where $ZC$ are the zero crossings, $EMG$ is the electromyographic signal, and n is a subset of integers related to the number of samples, $n$.

\subsubsection{Mean and median frequency }
\label{sec:Mean}

The mean and median frequencies, MNF and MDF, respectively, provide basic information about the signal spectrum and its changes as a function of time. They agree that the signal spectrum is symmetric about its center line, while their difference reflects a spectral bias. A tail in the high-frequency region implies higher MNF than MDF \cite{Merletti_2004}. Therefore, any reference spectral frequency can be used as an estimator of spectral compression. These parameters will be necessary to determine quantitative indices of the muscle activation pattern of the recorded signals. Namely, MNF represents the average (see equation \ref{eq:fmean}), and the MDF the signal's central position from the lowest to the highest values (see equation \ref{eq:fmedian}), \cite{Merletti_2004}.

\begin{equation}
f_{mean}=\frac{\int_{0}^{\frac{fm}{2}}f_{EMG}(f)\quad df}{\int_{0}^{\frac{fm}{2}}EMG(f) \quad df}
\label{eq:fmean}
\end{equation}

\begin{equation}
f_{median}=\frac{\int_{0}^{\frac{fm}{2}}(f-f_{mean})^k\ast EMG(f)\quad df}{\int_{0}^{\frac{fm}{2}}EMG(f)\quad df}
\label{eq:fmedian}
\end{equation}

Where $f_{mean}$ and $f_{median}$ represent the MNF and MDF, respectively, $f_{EMG}$ is the frequency of a specific electromyography channel, $f_m$ is the sampling frequency, and $k$ is the amount of data available.

\subsection{Coactivation index}
\label{sec:Coactivation}

The coactivation index is a measure of the activation of the agonist muscles, which exert an action in the same direction of movement, and antagonist muscles, which exert an action opposite to that of the other. To calculate muscle coactivation it is necessary to know the muscle activity during contraction \cite{Davidson_2013}. The percentage activation of each muscle during isometric contraction is calculated from the coactivation index, as follows:

\begin{equation}
CI=\frac{\int_{0}^{100}nEMGx \quad dt}{iEMG_{M1}+iEMG_{M2}+...+iEMG_{Mn}}
\label{eq:CI}
\end{equation}

Where $CI$ is the coactivation index, $nEMGx$ is the rectified linear envelope of the EMG signal of the muscle to be compared, and $iEMGMn$ is the integral of each muscle.

\subsubsection{Envelope with moving average}
\label{sec:Envelope}

The moving average is an arithmetic measure used to analyze a set of $N$ data in discrete time. This allows the creation of a series of averages, which can be simple or weighted. In the frequency domain, the moving average has a response of a low pass filter \cite{Roberts_2018}. For this purpose, the envelope with the moving average is calculated, as follows \cite{Proakis_2007}:

\begin{equation}
MA(N)=\frac{EMN(n)+ EMNG(n-1)+...+EMG(n-(N-1))}{N}
\label{eq:envelope}
\end{equation}

Where $MA$ is the moving average, $EMG$ is the electromyography signal of a channel, and $N$ is the number of samples. 

\section{Statistical methods}
\label{sec:Classification}

Classification section use methods consecutive artificial intelligence for data analysis, pattern classification, data mining and medical informatics through different Information visualization methods we can see the classification \cite{Tekin_2015}.

\subsection{Information visualization methods}

The visualization of the information allows understanding the behavior of the recorded data.

\subsubsection{Box plot}
\label{sec:Box}

\begin{figure}[h!]
    \centering
    \includegraphics[width=0.4\textwidth]{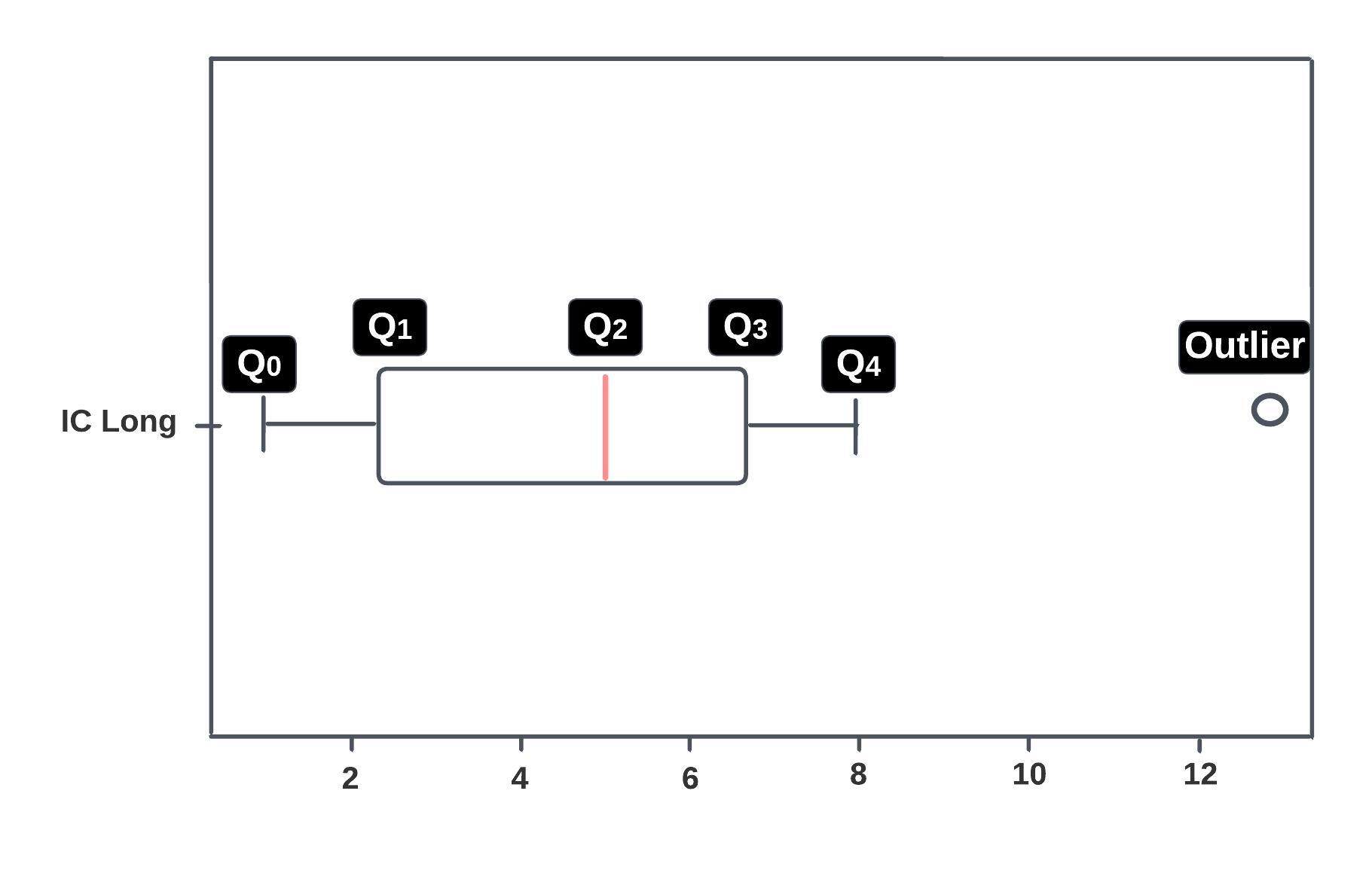}
    \caption{Box plot of the longissimus muscle coactivation index (IC Long). Where Q1 represents $25\%$ of the data, Q2 $50\%$ and Q3 $75\%$.}
    \label{fig:fig7}
\end{figure}

It is a type of graph that allows visualizing the locality, diffusion and asymmetry groups of a data set through its quartiles ($Q$). This diagram (see Figure \ref{fig:fig7}) is composed of a maximum (Q4) and a minimum ($Q0$) which is the upper and lower exclusion limit of the outliers, respectively, the median ($Q2$) and the upper ($Q3$) and lower ($Q1$) quartiles that represent the median of the upper and lower half \cite{Dutoit_1986}.

\subsubsection{Histogram}
\label{sec:Histogram}

The histogram is a method of data reduction by separating the information into bars, where each bar represents the amount in a frequency range for certain values. This diagram allows visualizing the distribution of the population \cite{Rufilanchas_2017}.

\subsection{Statistical tests}
\label{sec:Statistical}

Statistical tests are used to conclude a hypothesis using sample probability. It is composed of descriptive and inferential analysis. Inferential analysis aims to describe the population taking into account the data obtained from a sample. On the other hand, the descriptive analysis explains the trends of the sample. Three aspects must be taken into account when choosing the statistical test: the population, the sample size and the measurement scale. Statistical tests are divided into 2 sets: parametric and non-parametric \cite{Flores_2017}.

\subsubsection{Parametric}
\label{sec:headings}

Parametric tests can only be used if the data shows a normal distribution \cite{Flores_2017}.

\subsubsection*{Shapiro–Wilk test}
\label{sec:headings}

Shapiro–Wilk test is a parametric test based on a probability plot, in which the regression of the observations on the expected values of the hypothesized distribution is considered. The $W$ statistic represents the ratio of two estimates of the variance of a normal distribution (see equation \ref{eq:W}). This test has generally shown adequate results compared to the classic tests, but especially when working with short-tailed distributions and with a sample size of less than $30$, since it shows high variability when both the symmetry and the size are modified sample size of the distribution, especially between $20$ and $50$ participants. Its  statistic is defined as \cite{Pedrosa_2014}:

\begin{equation}
W=\frac{(\sum_{i=1}^{n}a_{i}x_{(i)} )^{2}}{\sum_{i=1}^{n}(x_{i}-\overline{x})^{2}}
\label{eq:W}
\end{equation}

where $x(i)$ is the number that occupies the $i-th$ position in the sample (with the sample ordered from the highest value). The variables $a_i$ are calculated, as follows:

\begin{equation}
(a_{1},...,a_{n})=\frac{m^{T}V^{-1}}{(m^{T}V^{-1}V^{-1}m)^{1/2}}
\label{eq:ai}
\end{equation}

where
\begin{equation}
m=(m_{1},...,m_{n})^{T}
\label{eq:mi}
\end{equation}

where $m_n$ are the mean values of the ordered statistic of independent and identically distributed random variables sampled from normal distributions and $V$ denotes the covariance matrix of that order statistic. The null hypothesis will be rejected if $W$ is too small. The value of $W$ can range from $0$ to $1$.

\textit{Interpretation}: being the null hypothesis that the population is normally distributed, if the $p$-value is less than $\alpha$ (significance level) then the null hypothesis is rejected, i.e., data do not have a normal distribution. If the $p$-value is greater than $\alpha$, it is concluded that this hypothesis cannot be rejected \cite{Shapiro_1965}.

\subsubsection*{D'Agostino test}
\label{sec:Agostino}

The objective of D'Agostino test is to establish whether or not the given sample comes from a normally distributed population (see equation \ref{eq:DA}). The test is based on transformations of the kurtosis and the skewness of the sample.

\begin{equation}
DA=\frac{\sum_{i=1}^{n}(i-(\frac{n+1}{2}))X_{i}}{n^{2}\sigma_{n}}
\label{eq:DA}
\end{equation}

where: $X_i$ indicates the data that appeared in place i in the sample, $X_i*$ are the ordered data in the sample, $n$: indicates the amount of data in the sample, and $\sigma_n$: is calculated using the equation \ref{eq:sigma}:

\begin{equation}
\sigma_{n}=\sqrt{\frac{\sum_{i=1}^{n}(X_{i}-\overline{X}_{n})^{2}}{n}}
\label{eq:sigma}
\end{equation}

In the table of the distribution of the D'Agostino statistic, for each level of significance, the calculated $D$ is compared; if it is less than the first member of the pair but greater than the second, then the null hypothesis of population normality is rejected \cite{Agostino_2017}.

\subsubsection{Nonparametric}
\label{sec:Nonparametric}

Non-parametric tests are based on random data, which seeks to identify behavior through different tests \cite{Flores_2017}.

\subsubsection*{Kolmogorov Smirnov test}
\label{sec:Kolmogorov}

The Kolmogorov test is a goodness-of-fit test, that is, the degree to which the observed distribution differs from another distribution (see equations \ref{eq:ks}). It is used when the number of data is small \cite{Berger_2014}.

\begin{equation}
KS=max_{x}|F_{1}(x)-F_{2}(x))|
\label{eq:ks}
\end{equation}

\begin{equation}
\notag if \quad KS > VC_\alpha \quad is \quad rejected \quad H_0
\label{eq:notag}
\end{equation}

\begin{equation}
\notag if \quad KS < VC_\alpha \quad is \quad not \quad rejected \quad H_0
\label{eq:notag_1}
\end{equation}

Where $KS$ is the test statistic, $max_x$ is the maximum difference between both distributions, the absolute value is used so the order of the operators does not alter the result.

\section{Summary}

sEMG has a great potential to support clinicians with objective and quantifiable information about individuals' health conditions affected by muscle disorders. For instance, lateral epicondylitis (LE) or tennis elbow in patients has been investigated by evaluating the coactivation of the forearm extensor muscles, given that it has been suggested that pain at the insertion of the forearm muscles at the epicondyle, affecting daily life activities (pressure effort, shaking hands or lifting a cup) \cite{Rojas_2019}. Other examples include measuring the resistance of the trunk muscles to evaluate low back pain \cite{Müller_2010, Wang_2019}. Given the complexity of the sEMG processing, this paper summarized relevant concepts that beginners in the field can use as a starting point in their studies. Specifically, the reader can get important insights into how to address a statistical analysis of sEMG signals in any application. For instance, at first, basic concepts of the muscle anatomy are indicated, to then introduce the reader to signal processing concepts, such as preprocessing signal, unique features of EMG, and statistical concepts that allow to analyze EMG data to be able to identify muscular patterns of diseases.

\section{Acknowledgments}
This work was supported by the Ministry of Science, Technology and Innovation (Minciencias) under Grant Cto. 489-2021 and the Universidad El Bosque under Grant PCI2019-10784. The authors thank Socrates Becerra for helping in the edition tasks.


\begin{thebibliography}{99}

\bibitem{Altimari_2012}
L.R. Altimari, J.L. Dantas, M. Bigliassi, T.F D. Kanthack, A.C. de Moraes, , and T. Abrao. Influence of Different Strategies of Treatment Muscle Contraction and Relaxation Phases on EMG Signal Processing and Analysis During Cyclic Exercise. \textit{Computational Intelligence in Electromyography Analysis - A Perspective on Current Applications and Future Challenges}. 2012. doi: 10.5772/50599

\bibitem{Berger_2014}
V.W. Berger and Y. Zhou. Kolmogorov–Smirnov Test: Overview. \textit{In Wiley StatsRef: Statistics Reference Online (eds N. Balakrishnan, T. Colton, B. Everitt, W. Piegorsch, F. Ruggeri and J.L. Teugels)}. 2014. doi: 10.1002/9781118445112.stat06558

\bibitem{Butterworth_1930}
S. Butterworth, (1930). On the Theory of Filter Amplifiers. Experimental Wireless \& The Wireless Engineer.

\bibitem{Correa_2016}
J. Correa, E. Morales, J. Huerta, J. González, C. Cárdenas. Sistema de Adquisición de Señales SEMG para la Detección de Fatiga Muscular. \textit{Rev. mex. ing. bioméd}, 37:1, 17-27, 2016.

\bibitem{Agostino_2017}
R.B. D’Agostino. Tests for the normal distribution. \textit{Goodness-of-fit techniques}. Routledge, 2017. 367-420.

\bibitem{Davidson_2013}
B. Davidson, D. Judd, A. Thomas, R. Mizner, D. Eckhoff, J. Lapsley. Muscle activation and coactivation during five-time-sit-to-stand movement in patients undergoing total knee arthroplasty. \textit{Journal of Electromyography and Kinesiology}, 23(6):1485-1493, 2013.

\bibitem{Drost_2013}
G. Drost, D.F. Stegeman, B.G.M. van Engelen, M.J. Zwarts, Clinical applications of high-density surface EMG: A systematic review. \textit{Journal of Electromyography and Kinesiology}, 16(6):586-602, 2006, doi: 10.1016/j.jelekin.2006.09.005

\bibitem{Dutoit_1986}
S.H.C. du Toit, A.G.W. Steyn and R.H. Stumpf. \textit{Graphical Exploratory Data Analysis}. Springer-Verlag, New York, 1986.

\bibitem{Flores_2017}
E. Flores, M. Miranda, M. Villasís. El protocolo de investigación VI: cómo elegir la prueba estadística adecuada. Estadística inferencial. \textit{Rev. alerg. Méx.}, 64(3), 2017.

\bibitem{Gomez_2009}
C. Gomez, C. Bolaños, B. Minaya, H. Fogaca. Mecanismos implicados en la fatiga aguda. \textit{Revista Internacional de Medicina y Ciencias de la Actividad Física y el Deporte}, 10:40, 537-555, 2009.

\bibitem{Guerrero_2014}
F. Guerrero, M. Haberman, E. Spinelli. Multichannel Biopotential Acquisition System. \textit{Revista Ingeniería Biomédica}, 8(15):18–26, 2014.

\bibitem{Guyton_2006}
A.C. Guyton, and J.E. Hall. Textbook of Medical Physiology. 2006. 11th Edition, Elsevier Saunders, Amsterdam. 

\bibitem{Kim_2014}
D. Kim, Y. Cho, J. Ryu, Real-time locomotion mode recognition employing correlation feature analysis using emg pattern. \textit{ETRI Journal}, 36(1):99–105, 2014. 

\bibitem{Konard_2005}
Konard P. The ABC of EMG: A Practical Introduction to Kinesiological Electromyography. Noraxon Inc. USA, 2006. 

\bibitem{Macintosh_2002}
B. Macintosh, E. Dilson. What is fatigue?. \textit{Can J. Appl Physiol.} 27(1): 42-55, 2002.

\bibitem{Mañanas_1999}
M. Mañanas. Análisis de la actividad muscular respiratoria mediante técnicas temporales, frecuenciales y estadísticas. Dissertation. Universidad Politecnica de Catalunya. 1999.

\bibitem{Melo_2012}
W.C. Melo, E.B.L. Filho, W.S.S. Junior. Electromyographic signal compression based on preprocessing techniques. \textit{2012 Annual International Conference of the IEEE Engineering in Medicine and Biology Society}. 2012. doi: 10.1109/embc.2012.6347216

\bibitem{Merletti_2004}
R. Merletti, P. Parker. Electromiography, Physiology, Engineering, and Noninvasive Applications. 2004. United States, New Jersey. IEEE. 

\bibitem{Merletti_2010}
R. Merletti, A.  Botter, C.  Cescon, M.A.  Minetto and T.M.M.  Vieira. Advances in Surface EMG: Recent Progress in Clinical Research Applications. \textit{
Critical Reviews: in Biomedical Engineering}. 38(4):347-379, 2010.

\bibitem{Merletti_2016}
R. Merletti and D. Farina,  (Eds.). Surface electromyography: physiology, engineering, and applications. 2016. John Wiley \& Sons.

\bibitem{Merletti_2020}
R. Merletti, G.L. Cerone. Tutorial. Surface EMG detection, conditioning and pre-processing: Best practices. \textit{Journal of Electromyography and Kinesiology}, 54, 2020. doi: 10.1016/j.jelekin.2020.102440.

\bibitem{Mohr_2018}
M. Mohr, T. Schön, V. Tscharner, M. Nigg. Intermuscular coherence between surface EMG signals is higher for monopolar compared to bipolar electrode configurations. \textit{Frontiers in Physiology}, 1–14. 2018.

\bibitem{Mora_2013}
Mora, I. (2013). Detección de crisis epilépticas a partir de señales EEG mediante índices basados en el algoritmo de Lempel-Ziv (Bachelor's thesis, Universitat Politècnica de Catalunya).

\bibitem{Mordhorst_2015}
M. Mordhorst, T. Heidlauf, O. and Röhrle. Predicting electromyographic signals under realistic conditions using a multiscale chemo–electro–mechanical finite element model. \textit{Interface Focus}, 5(2):20140076. 2015 doi:10.1098/rsfs.2014.0076.

\bibitem{Müller_2010}
R. Müller, K. Strässle, B. Wirth. Isometric back muscle endurance: an EMG study on the criterion validity of the Ito test. \textit{J Electromyogr Kinesiol}. 2010 20(5):845-50, 2010. doi: 10.1016/j.jelekin.2010.04.004. 

\bibitem{Pedrosa_2014}
I. Pedrosa, J. Juarros-Basterretxea, A. Robles-Fernández, J. Basteiro and E. García-Cueto Pruebas de bondad de ajuste en distribuciones simétricas, ¿qué estadístico utilizar?. \textit{Universitas Psychologica}, 14(1):245-254, 2015. doi: 10.11144/Javeriana.upsy13-5.pbad

\bibitem{Phinyomark_2011}
A. Phinyomark, C. Limsakul and P. Phukpattaranont. Application of wavelet analysis in EMG feature extraction for pattern classification. \textit{Measurement Science Review}, 11(2):45-52. 2011. doi:10.2478/v10048-011-0009-y

\bibitem{Pilkar_2020}
R. Pilkar, K. Momeni, A. Ramanujam, M. Ravi, E. Garbarini and G.F. Forrest. Use of Surface EMG in Clinical Rehabilitation of Individuals With SCI: Barriers and Future Considerations. \textit{Frontiers in Neurology}, 11, 2020. doi: 10.3389/fneur.2020.578559, 

\bibitem{Proakis_2007}
J. Proakis, D. Manolakis. Digital Signal Processing Principles, Algorithms, and Applications. 2007. Pearson prentice hall, Fourth Edition.

\bibitem{Quinayas_2015}
C. Quinayás, C. Gaviria. Sistema de identificación de intención de movimiento para el control mioeléctrico de una prótesis de mano robótica. \textit{Ingeniería y Universidad}, 19(1):27-50, 2015. doi: 10.11144/Javeriana.iyu19-1.siim

\bibitem{Quiroz_207}
G. Quiroz. Laboratorio Analogico. Capitulo II, Teoría de filtros, Bachelor's Thesis,  Universidad de las Américas Puebla, Escuela de Ingeniería y Ciencias, Departamento de Computación, Electrónica y Mecatrónica. 2007. 

\bibitem{Roberts_2018}
M. Roberts. \textit{Signals and Systems: Analysis Using Transform Methods \& MATLAB}. 2018.

\bibitem{Rojas_2019}
M. Rojas-Martínez, J.F. Alonso, M. Jordanić, M.Á. Mañanas and J. Chaler. Analysis of Muscle Load-Sharing in Patients With Lateral Epicondylitis During Endurance Isokinetic Contractions Using Non-linear Prediction, \textit{ Front. Physiol.}, 10:1185, 2019. doi: 10.3389/fphys.2019.01185

\bibitem{Romo_2007}
H. Romo, J. Realpe, P. Jojoa. Análisis de Señales EMG Superficiales y su Aplicación en Control de Prótesis de Mano. \textit{Revista Avances en Sistemas e Informática}, 4(1):127-136, 2007.

\bibitem{Rufilanchas_2017}
D. Rufilanchas. On the origin of Karl Pearson’s term “histogram”. \textit{Estadística Española}, 59(192):29-35, 2017.

\bibitem{Salgado_2015}
J. Salgado, L. Córdoba, M. Dussan. Medidor de actividad eléctrica muscular cuatro canales electromiógrafo (EMG) inalámbrico. \textit{Revista Ingeniería y Región}. 2015;13(1):201-213, 2015.

\bibitem{Shapiro_1965}
S.S. Shapiro, M. B. Wilk. An analysis of variance test for normality (complete samples). \textit{Biometrika}. 52(3-4):591-611. 1995. doi: 10.2307/2333709.

\bibitem{Tekin_2015}
T. Tekin Erguzel, C. Tas, M. Cebi. A wrapper-based approach for feature selection and classification of major depressive disorder–bipolar disorders. \textit{Computers in Biology and Medicine}. 64:127–137, 2015. doi:10.1016/j.compbiomed.2015.06.

\bibitem{Thompson_2010}
J. C. Thompson. Netter's Concise Orthopaedic Anatomy. Elsevier. 2010

\bibitem{Wang_2019}
N. Wang, Z. Zhang, J. Xiao and L. Cui. DeepLap: A Deep Learning based Non-Specific Low Back Pain Symptomatic Muscles Recognition System. \textit{In 2019 16th Annual IEEE International Conference on Sensing, Communication, and Networking (SECON)}.

\bibitem{Welch_1967}
P. Welch. The use of fast Fourier transform for the estimation of power spectra: A method based on time averaging over short, modified periodograms. \textit{IEEE Transactions on Audio and Electroacoustics}, 15(2):70-73, 1967. doi: 10.1109/TAU.1967.1161901.

\bibitem{Yousefi_2014}
J. Yousefi, A. Hamilton-Wright. Characterizing EMG data using machine-learning tools. \textit{Computers in Biology and Medicine}, 51:1-13, 2014.
doi: 10.1016/j.compbiomed.2014.04.018.

\end{thebibliography}
\end{document}